# Talbot-Lau interferometry with fullerenes: Sensitivity to inertial forces and vibrational dephasing

A. Stibor, K. Hornberger, L. Hackermüller, A. Zeilinger and M. Arndt
e-mail: markus.arndt@univie.ac.at

**Topic:** Fundamental Problems

**Abstract**
We discuss matter wave experiments in a near-field interferometer and focus on dephasing phenomena due to inertial forces. Their presence may result in a significant reduction of the observed interference contrast, even though they do not lead to genuine decoherence. We provide quantitative estimates for the most important effects and demonstrate experimentally the strong influence of acoustic vibrations. Since the effects of inertial forces get increasingly important for the interferometry with more massive particles they have to be identified and compensated in future experiments.

*We dedicate this article to Herbert Walther at the occasion of his 70$^{th}$ birthday. Herbert Walther inspired already more than a decade ago a discussion on complimentarity in interference experiments which is still of relevance today.*

# 1 Interferometry with large molecules

In recent years, there has been an increasing interest in the coherent manipulation of molecules [1-5] and there were significant advances on the way to extend matter wave interferometry to larger cold clusters [4] and hot macromolecules [5].

A major challenge in such experiments is to find sufficiently intense sources, a scheme to prepare the required coherence and a detector that can resolve the small effects related to the tiny size of the de Broglie wavelength. Typical molecular beams have velocities between 100 m/s and 2000 m/s with corresponding de Broglie wavelengths of several picometers and longitudinal and transverse coherence lengths of the same order of magnitude.

Far-field diffraction therefore requires a tight collimation of the beam, which imposes a severe limit on the signal strength. It also requires the usage of very tiny diffraction structures. Current state of the art technology allows the fabrication of gratings with periods down to 100 nm and open slit widths of about 50 nm [6]. But even though it can be expected that five to ten times smaller structures may be manufactured in the future, these nano-elements will probably be of rather limited use because they will then be as small as the large diffracted objects themselves.

One way around the "size problem" is the three-grating near-field interferometer, which can act as a lens-less imaging device with high signal throughput [7]. In a near-field setup the required grating period only scales with the square root of the de Broglie wavelength, which is superior to far-field diffraction where both the length of the apparatus and the grating period scale linearly in the wavelength. Also, a near-field interferometer of the Talbot-Lau type accepts beams of rather poor initial transverse coherence. This feature allows to significantly increase the count rates in the experiment in comparison to standard far-field setups.

It was recently shown, that a Talbot-Lau interferometer works indeed very efficiently for $C_{70}$ molecules and that the integration time for one interferogram could be shortened by more than two orders of magnitude [8] as compared to former far-field studies in the same machine. The Talbot-Lau configuration was therefore also a prerequisite for the successful demonstration of the wave-particle duality of more massive and larger objects like the fluorofullerenes and the porphyrins [9] since typical beam intensities and detection efficiencies for these species prohibited far-field experiments.

Near-field interferometry based on gratings with a large period has the additional property of being very sensitive to inertial forces, such as those due to the rotation and the gravitational acceleration of the earth. Of course, this implies also a high sensitivity to perturbing acoustic noise and floor vibrations. Our present study therefore explores the influence of inertial forces and vibrational dephasing to deduce the stability requirements for future experiments.

# 2 The Talbot-Lau effect

The basic setup of our interferometer has already been described previously [8] and it will therefore only be summarized (cf. Fig. 1): Molecules from a thermal source are evaporated into a vacuum chamber. The velocity selection of the molecular beam is done using three narrow slits, which select a well-defined free-flight parabola and thus a narrow velocity class from the beam. The base pressure is set to below 5x10$^{-8}$ mbar in order to avoid the influence of collisional decoherence that had been observed in earlier experiments [10]. The molecules pass a symmetric Talbot Lau interferometer, which consists of three identical gold gratings with a grating constant of 990 nm, which are separated by $L$=0.38 m, respectively, i.e. by the Talbot length $L_T = d^2/\lambda_{dB}$ for a mean de Broglie wavelength of $\lambda_{dB}$ = 2.58 pm. Those molecules, which pass the third grating, enter the thermal ionization detector [11].

The first grating prepares the molecular transverse coherence out of the originally uncollimated and therefore spatially incoherent beam. Diffraction at the second grating then produces a molecular density pattern at the location of the third grating. Whenever the molecular pattern and the third grating mask fall in line, the transmission is high. When the third grating is shifted by half a grating period, the total transmitted signal reaches its minimum. Plotting the molecule counts as a function of the displacement of the third grating thus leads to the observation of a fringe pattern, which is the signature of molecular interference. The full circles in Fig. 2 clearly show that fringes of high visibility can be obtained in the experiment.

The pure *Talbot* effect, which applies to *plane-parallel* illumination of a single grating, produces coherent self-images of a diffraction grating, with periodic recurrences in multiples of the Talbot length. In the symmetric *Talbot-Lau* configuration the first grating illuminates the second one with *cylindrical* wavelets, which leads to interference patterns in integer multiples of the Talbot distance [12,13]. If we include the attractive van der Waals interaction between the grating walls and the molecules, the imaging distance is shifted but we still observe a recurrence of the molecular interference pattern when the grating separation is varied.

Fig. 3 shows the formally equivalent case where the molecular de Broglie wavelength, i.e. the molecular velocity, is varied. We see a very good agreement between quantum expectation (solid line) and experiment (full circles) for short de Broglie wavelengths ($\lambda_{dB}$ ~ 2.5 pm), but we also notice a clear contrast reduction in the experiments for slow molecules ($\lambda_{dB}$ ~ 5 pm).

Essentially all fullerenes leave the oven in their electronic ground state, which is diamagnetic, has no electric dipole moment, and which is known to radiate infrared quanta only at very low rates and at very long wavelengths [14,15]. It thus seems that there is no mechanism which could lead to any irreversible coupling to the environment. Therefore, we do not expect genuine decoherence as described in [16] and references therein.

However, there are several mechanisms which may lead to a velocity dependent shift of the interference pattern. The velocity distribution of real beams is always finite. The observed interference contrast will therefore be reduced if the detection does not resolve the different velocities, because many interference patterns will be shifted with respect to each other and tend to average.

## 3  Inertial forces in the Talbot-Lau setup

The appearance of inertial phase shifts can be very important in matter wave interferometry. In fact, atom interferometers are currently among the best devices to measure accelerations [17,18,19].

Inertial forces arise if the frame of reference, defined by the interference apparatus, is not an inertial system. The two best known effects are the gravitational and the Coriolis acceleration due to the mass and the rotation of the earth. In addition to that, the coupled gratings may oscillate about their position at rest. Here we focus mainly on externally excited vibrations which drive the gratings at equal frequencies but with different amplitudes.

The effect of inertial forces on the interference fringe pattern is calculated most conveniently by considering the dynamics in a co-moving reference frame, in which the gratings are fixed. The prediction for the interference pattern is then obtained by propagating the Wigner function of the matter wave beam through the apparatus [12], and taking into account that the canonical momentum differs from the kinetic momentum due to the presence of the inertial forces. An advantage of the Wigner formulation is that the corrections due to a finite longitudinal coherence length and due to the grating interactions are automatically taken into account in the calculation. As a result, we find in all cases below that the resulting fringe pattern is unchanged, but shifted laterally by a distance $\Delta x$ which depends on the velocity of the particle. It is worth noting that the same displacement is found in the corresponding description of classical particles forming a Moiré pattern [20]. This can be related to the fact that the inertial forces are at most linear in position and momentum so that the free evolutions of the Wigner functions and of the classical phase space density are equal. Therefore, it is sufficient to follow the trajectory of classical particles to obtain the observed shift of the interference fringes.

### 3.1 Rotation of the earth

We start by considering the effect of the Coriolis force, assuming that all gratings are vertically aligned and that there are no vibrational perturbations. Let $\Omega_0$ be the parallel component of the angular velocity vector with respect to the grating bars. Then the fringe pattern shifts by

$$\Delta x = \frac{2\Omega_0 L^2}{v_z} = \Omega_0 L \tau \quad (1)$$

as compared to the case without acceleration. Here $L$ is the distance between two gratings and $v_z$ is the molecular velocity. The fringe shift increases linearly with the time of flight $\tau = 2L/v_z$ in an apparatus of fixed length. The displacement (1) neither depends on $\hbar$ nor on the mass $m$ of the particle. This is consistent with other formulations of the Sagnac phase [21] $\Phi_{\text{Sagnac}} = 2m\vec{\Omega}_0 \vec{A}/\hbar$, as often used in neutron [22] and atom interferometry [18]. Indeed, if we set the interferometer to the first Talbot order, $L = L_T$, we find $\Delta x / d = \Phi_{\text{Sagnac}} / 2\pi$ if $A = dL$ is the area enclosed by neighboring interfering paths. For the N-th Talbot order, $L = NL_T$, the corresponding area is $A = NdL$. A related argument holds in a far-field Mach-Zehnder interferometer, as can be easily verified by noting that the interferometer surface depends on the diffraction angle behind the first grating.

For our experiments we calculate a Coriolis shift of $\Delta x = 80$ nm, with $\Omega = 5.55 \times 10^{-5}$ rad/s (48° northern latitude), L=0.38 m and $v_z = 200$ m/sec. This shift is certainly larger than the detector resolution but it only effects the fringe visibility, if the molecules have a broad velocity spread. If we assume a Gaussian velocity distribution with standard deviation $\sigma_v$ we find that the visibility $V$ of a sinusoidal fringe pattern is reduced by the factor

$$R_C = \exp\left(-8\left[\pi \frac{\Omega_0 L^2 \sigma_v}{d\, v_z^2}\right]^2\right) \quad (2)$$

For $d$=990 nm and $\sigma_v = 0.1 v_z$ the visibility remains almost unaffected by $R_C$=99.8%, for 200 m/sec, and $R_C$=99.5% for 100 m/sec.

However, for larger masses considerably lower velocities will be needed to keep the de Broglie length and therefore also the interferometer length acceptable. For example, if we replaced our fullerenes by a ten times more massive object, such as a protein or nanocrystal with a mass of 10,000 amu, this object would already have to be slowed or cooled to a velocity of $v_z = 10$ m/s. This corresponds to the most probable velocity of a thermal cloud at T~60 K. If we also kept the same velocity spread as before ($\sigma_v = 0.1 v_z$), the visibility would already be reduced by a factor of $R_C$=59.6%.

In order to see how $R_c$ scales with the molecular mass, we now fix the source properties $v_z$ and $\sigma_v$ and we assume that the Talbot Lau interferometer will always be operated in the lowest Talbot order. If we require that the visibility may not be reduced by more than a factor $1/e$ we can distinguish three cases:

First, if the grating period $d$ is kept constant, the interferometer length increases linearly with the mass $m$, since $L_T = d^2/\lambda_{dB} = d^2 m v/h$. In this case the particle mass is bounded by

$$m < \hbar \left(\frac{\sqrt{2}\pi}{\Omega_0 d^3 \sigma_v}\right)^{1/2} \quad (3).$$

Second, if the length of the interferometer is kept constant the grating period $d$ scales with the inverse square root. In that case the particle mass is bounded by

$$m < \frac{\hbar v_z^3}{4\pi \Omega_0^2 L^3 \sigma_v^2} \quad (4).$$

The mass limit can thus be pushed almost arbitrarily by improving on the velocity spread of the molecules. In the near future we will be bound by both, $d \geq 100$ nm and $L \leq 1$m due to practical limitations. In this third case (2) yields a requirement for the velocity selection which is independent of the mass,

$$\frac{\sigma_v}{v_z^2} < \frac{d_{min}}{\sqrt{8\pi}\Omega L_{max}^2} \cong 2\times 10^{-4}\,\frac{s}{m}.$$

Finally, we note that the Sagnac shift can be completely avoided by orienting the interferometer such that its surface vector is perpendicular to the rotation axis of the earth. Generally, this would require a vertical alignment of the molecular beam and a specific orientation of the gratings with respect to the north pole.

### 3.2 Gravitational acceleration

The influence of gravity enters the reasoning if two interfering paths in the apparatus experience different gravitational potentials, that is, if the grating bars are not aligned perfectly parallel to the vertical direction. Let $\theta_G$ be the angle between the grating bars and the direction of the gravitational centre. Then the Talbot-Lau fringes shift by

$$\Delta x = g\sin\theta_G \frac{L^2}{v_z^2} \quad (5)$$

with respect to the case without acceleration. This is half of the deflection a classical particle experiences due to the acceleration $a = g\sin\theta_G$, and it is again equal to the displacement of the Moiré fringes in the corresponding classical dynamics [19].

The Coriolis fringe shift (1) is formally equivalent to (5) if we simply replace the gravitational acceleration $a_G = g\sin\theta_G$ by the Coriolis acceleration $a_C = 2v_z\Omega_0$. Therefore, equations (2) – (4) also describe the effect of earth's gravity if $\Omega_0$ is replaced by $g\sin\theta_G/(2v_z)$. In particular,

$$R_G = \exp\left(-2\left[\pi\frac{g\sin\theta_G L^2 \sigma_v}{d\, v_z^3}\right]^2\right). \quad (6)$$

We took care to pre-align our gratings to better than $\theta_G = 1\times 10^{-3}$ rad. With the parameters given above and $g$=9.81 m/s² this implies that the fringe pattern may shift up to 36 nm for molecules at 200 m/s, and already by 144 nm at 100 m/s. For fast molecules the visibility reduction due to gravity is therefore of no concern. However, molecules at 100 m/s require a four times better grating alignment. Although this can be done by sequentially rotating the grating mounts to optimize the fringe visibility, it can not be excluded that the optical table as a whole exhibits slow tilting drifts which are of the order of 1 mrad.

But again, the constant fringe shift itself does not contribute to the loss of visibility, and with a velocity selection of $\sigma_v/v_z = 10\%$ gravity will only reduce the contrast by less than one percent, even for molecules at 100 m/s. However, unlike the Coriolis force, the gravitational acceleration is not reduced as the particle velocity decreases. Therefore, it is bound to become rather relevant for high masses at low velocities. This can also be seem by comparing Eq. (4) and Eq. (6).

# 4  Grating vibrations

Another important source of dephasing are vibrational perturbations, which give rise to time dependent inertial forces. If the oscillation periods are much longer than the passage time through the interferometer the corresponding accelerations may be taken as constant during the time of flight and Eqs. (1) and (5) may still be used. However, the situation gets more complicated if we consider oscillation frequencies which are comparable to the inverse time of flight or even larger. Typically, these frequencies are in the acoustic regime between 50 and 1000 Hz which are in fact dominating the vibrational spectrum of in the experiment due to the presence of vacuum pumps, laser cooling circuits and other experimental equipment on the table.

One can classify the vibrations of single gratings according to their direction of motion as (a) 'transverse shifting', where the gratings move along the grating k-vector, (b) 'forward shifting', which describes a change in the grating separation, (c) 'upward shifting' vibrations, which lift and lower the gratings, (d) 'rolling' oscillations which describe the rotation of the grating around the molecuar beam axis, (e) 'downward tilting', where the tip of the grating tilts downwards and finally (f) 'yawing', where the gratings are rotated around a vertical axis, parallel to the grating lines.

Most of these motions can be immediately eliminated again from the list of detrimental perturbations: It can be shown that 'forward shifting oscillations' (b) are only relevant for amplitudes larger than 50μm, which simply are not excited – even in the worst case. 'Upward shifting' (c) is irrelevant since the gratings are vertically oriented and translation invariant over the relevant distances. 'Downward tilting' (e) as well as 'yawing' (f) are actually variants of case (b), where the forward shift depends on the height or lateral position on the grating, respectively.  They can therefore also be neglected. 'Rolling oscillations' (d) can be reduced to height dependent transverse shifts (a) and enter only to heigher order.

We are therefore left with 'transverse shifts' (a), which can of course be detrimental even on the sub-micrometer level. The transverse shifts segregate into coupled mode oscillations, most importantly the 'fixed pendulum' and the 'torsion pendulum' mode, as well as into independent grating oscillations. Since these vibrations reduce the visibility even for monochromatic beams we disregard the width in the velocity distribution. Our results may then easily be extended by averaging over the real velocity distribution.

## 4.1  Fixed pendulum oscillations

The basic oscillation mode affecting the fringe visibility is a common horizontal motion of the gratings perpendicular to the particle beam. The corresponding fringe shift is

$$\Delta x = A[\sin(\varphi_0) - 2\sin(\varphi_0 - 2\pi f L / v_z) + \sin(\varphi_0 - 4\pi f L / v_z)] \quad (7)$$

with $A$ the oscillation amplitude, $f$ the frequency and $\varphi_0$ the initial phase position of the gratings.

To obtain the observed interference contrast the shifted patterns must be averaged over the phase position $\varphi_0$ which is distributed uniformly, since the time of arrival is uncorrelated with the grating motion. This way one finds the reduction factor

$$R_{\mathrm{P}} = \left| J_0\left( 8\pi \frac{A}{d} \sin^2\left( \pi \frac{fL}{v_z} \right) \right) \right| \quad (8)$$

with $J_0$ the zero-order Bessel function. Figure 4 shows the reduction factor (8) for the parameters of our experiments (L=0.38m, $v_z$=200 m/s) and *A/d*=0.5 as a function of the frequency *f*. As expected, the visibility is not reduced if, the grating performs an integer number of sinusoidal motions during the travel of the molecule between two gratings. Equally, the contrast reduction is strongest for an odd number of half-sinusoidal motions.

### 4.2 Torsional pendulum oscillations

The second harmonic oscillation to be considered is the general torsional pendulum motion about a fixed point. In this case the angular velocity oscillates with frequency *f* so that $\Omega(t) = \Omega_0 \sin(\varphi_0 + 2\pi f t)$. We take $z_0$ to be the longitudinal position of the first grating with respect to the axis of rotation so that for $z_0 = 0$, $z_0 = -L$, and $z_0 = -2L$ the rotation is about the first, the second, and the third grating, respectively. We find

$$\Delta x = \frac{\Omega_0}{2\pi f} \left[ z_0 \cos(\varphi_0) - 2(z_0 + L)\cos(\varphi_0 - 2\pi f L / v_z) + (z_0 + 2L)\cos(\varphi_0 - 4\pi f L / v_z) \right]. \quad (9)$$

Note that the term of the k$^{\mathrm{th}}$ grating (k=1,2,3) vanishes if the axis of rotation lies in the plane of this grating. Like above, one has to average over the initial phase position $\varphi_0$ in order to obtain the visibility reduction and we obtain the reduction factor

$$R_{\mathrm{T}} = \left| J_0\left( \sqrt{8} \frac{\Omega_0 L}{fd} \sin\left( \pi \frac{fL}{v_z} \right) \sqrt{1 + \left(1 + \frac{z_0}{L}\right)^2 - \frac{z_0}{L}\left(2 + \frac{z_0}{L}\right)\cos\left(2\pi \frac{fL}{v_z}\right)} \right) \right|. \quad (10)$$

In Figure 5 we plot $R_{\mathrm{T}}$ with our experimental parameters as a function of the frequency *f* and as a function of the pivot point position in units of *L*. For small frequencies, $R_{\mathrm{T}}$ is independent of $z_0$ and approaching

$$R_{\mathrm{T}}(f \to 0) = J_0\left( 4\pi \frac{\Omega_0 L^2}{dv_z} \right).$$

### 4.3 Independent grating vibrations

Finally, we consider the case that the three gratings exhibit independent shifting oscillations, without a fixed phase relation. In this case one has to average over the fringe shifts, which correspond to the various grating configurations, by taking into account the probability distribution of the grating positions.
We assume that the three gratings oscillate harmonically with the amplitudes $A_1$, $A_2$, and $A_3$. Then the average yields

$$R_{\mathrm{I}} = \left| J_0\left(2\pi \frac{A_1}{d}\right) J_0\left(4\pi \frac{A_2}{d}\right) J_0\left(2\pi \frac{A_3}{d}\right) \right|. \quad (11)$$

This reduction is independent of the oscillation frequencies because the probability distribution does not depend on them. Note that the second factor in (11), which belongs to the second grating, has an argument that is twice as large. This also holds in the classical case and can there be simply related to the theorem of intersecting lines.

One may also consider position distributions of the gratings which are not due to a harmonic oscillation. An obvious choice is a Gaussian distribution, say, due to the Brownian motion in a harmonic potential. In this case the visibility reduction reads $R = \exp(-2\pi^2(\sigma_{A1}^2 + 4\sigma_{A2}^2 + \sigma_{A3}^2)/d^2)$, if the standard deviations are given by $\sigma_{A1}, \sigma_{A2}$, and $\sigma_{A3}$, respectively [23].

## 5 Experimental exploration of vibrational noise

In order to understand and quantify the influence of acoustic noise in our experiment we rigidly connected accelerometers (Bruel&Kjaer, Type 8318) both to the outside of a horizontal and a vertical flange of the vacuum apparatus. The output of these sensors was read and Fourier-analyzed by a fast oscilloscope (LeCroy Waverunner LT374M). With the specified sensor calibration factor $k$=316 mV/(m/s$^2$), we determine both the vibrational acceleration $a$ and the amplitude $x$ from the frequency dependent voltage $U$ and find $x = a/\omega^2 = U/k\omega^2$. We typically obtain amplitudes up to several 100 nanometers. where the largest amplitudes are found at frequencies below 100 Hz, with a significant peak around 50 and 100 Hz, where the surrounding roughing pumps have their largest acoustic contribution.

A first and crucial step to eliminate perturbations was to inflate the pneumatically levitated optical table (Melles Griot, SuperDamp) to which the whole vacuum machine was rigidly bolted. Inflating the optical table reduces characteristic vibrations at 50 Hz and 100 Hz by about 20 dB. The result is dramatic and can be seen by comparing the interference curves in Fig. 2. Without the vibration isolation the interference visibility was reduced by at least a factor of four. And on many days, the overall laboratory floor noise was so influential, that no interference could be observed at all without the efficient decoupling by the air damped table feet.

The closed-circle water cooling of the Argon ion laser, represented a second serious low-frequency noise source. It could be isolated from the experiment on the same table through carefully selected pieces of rubber. It turned out that standard erasers, as obtained from a stationery shop, out-performed other tested laboratory materials (viton, teflon) in damping efficiency. The use of the erasers was imperative to increase the best observed overall interference fringe contrast from 39% to 50%.

Although both noise isolation methods were required and successful in obtaining the theoretically expected interference visibility for fast molecules (200 m/s, $\lambda_{dB}$=2.5 pm, Fig. 3), they still did not suffice to fully restore the expected interference contrast at 100 m/s ($\lambda_{dB}$=5 pm).

In order to better demonstrate the influence of acoustic noise on the interference pattern it was easier to add than to suppress selected frequencies. This was done by mounting a PC loudspeaker to a horizontal flange on the vacuum chamber, laterally separated by approximately 30 cm from the horizontal accelerometer. The loudspeaker was driven by a sine-function generator. The accelerometers clearly identified this contribution, and higher harmonics of the chosen frequency were only excited at high vibration amplitudes. We then recorded interference patters for various acoustic frequencies and kept the absolute displacement amplitude at the location of the accelerometer constant to nominally 15 nm only. When we plot the observed fringe visibilities as a function of the applied frequency we find a multiply peaked curve with a dramatic loss of interference contrast at several frequencies, as shown in Fig. 6. The overall rich peak structure is not unexpected, given our insights of Sec. 4. However, the peaks are not equidistant in frequency and a direct comparison between theory and experiment is hampered by the unavoidable mechanical resonances both of the vacuum chamber and the interferometer mounts. They enhance the detrimental dephasing at some frequencies, in particular around 100 Hz and 130 Hz and suppress them at others. Although a quantitative analysis is thus rendered difficult the experiments certainly show that even very small mechanical perturbations on one part of the experiment can cause a large effect on the overall performance of the interferometer. And it will be a significant challenge to identify and exclude all possible sources of acoustic dephasing in future experiments with larger clusters and molecules.

## 6    Conclusion

We have shown that a Talbot Lau interferometer is not only a suitable device for investigating the wave-particle duality of large molecules, but it is also a very sensitive to inertial accelerations, be it due to rotational, gravitational or acoustic perturbations.

Our present study shows that the rotation of the earth is not yet a limiting factor for interference with particles in the 1000 amu range. But the Coriolis force has to be taken into account in near-field interferometry with large proteins, unless the molecular velocity spread can be reduced to below $\sigma_v / v_z \approx 1\%$.

The suppression of the gravitational phase shift is already a major concern in the design of our current Talbot interferometer and it is compensated by an alignment of the grating bars with respect to the vector of gravity to better than 1 mrad.

The third source of dephasing in the present interferometer is acoustic noise. Extending previous studies of these effects [23], we predict the contrast reducing influence of the basic acoustically excited vibration modes of the Talbot Lau interferometer. Experimentally, we were able to fully compensate all perturbations to such a level, that interferograms of 'fast' molecules (200 m/s) reached the theoretically expected visibilities. In order to achieve that, it turned out to be crucial to rigidly mount the whole vacuum machine on an air-damped optical table and in addition to isolate noise sources on the table. However, for slow molecules (100 m/s) the observed contrast is still about 30 % lower than expected. Interferometers with multiple gratings, more massive and slower molecules will therefore require additional care in avoiding mechanical vibrations. This can in principle be provided by passive and active methods. Novel solid state lasers require much less cooling water and magnetically levitating turbo molecular pumps or large ion getter pumps will significantly reduce the acoustic noise. The next generation of our experiments will be built in a new laboratory on an isolated concrete basement three floors below the current setup, to minimize the influence of building vibrations.

It is also worth noting that most of the perturbing influences could be drastically reduced and compensated in a future satellite based experiment – even though the costs for such an effort would not be justified at present.

We conclude with an expample for a proposed experiment with insulin, which has a mass of about 5700 amu, and which would traverse a Talbot Lau interferometer with L=0.4 m, d=257 nm, v=300 m/s and σ=30 m/s, θ =0.001 rad and $\Omega = 5.55 \times 10^{-5}$ rad/s as before. Assuming that the vibration amplitudes can be controlled to within A=10 nm, we predict the following contrast reduction factors: $R_C$=0.99, $R_G$= 0.999, $R_P$ > 0.75 for all frequencies and $R_I$ =0.91. An estimation of $R_T$ would require a precise knowledge of the pivot point, but it is reasonable to assume that the perturbing influence of torsion pendulum oscillations will generally not exceed that of the other modes. This shows, that vibrations will still be represent the most influential limitation of our interference design. But a reduction of all vibration amplitudes to below 5 nm will again render the interference contrast almost perfect.

**Acknowledgments:**
Our experiments are funded by the Austrian Science Foundation (FWF) within the projects START Y177 and SFB F1505 as well as by the European Commission under contracts HPRN-CT-2000-00125 and HPRN-CT-2002-00309. K.H. gratefully acknowledges support by the Emmy Noether program of the Deutsche Forschungs-gemeinschaft.

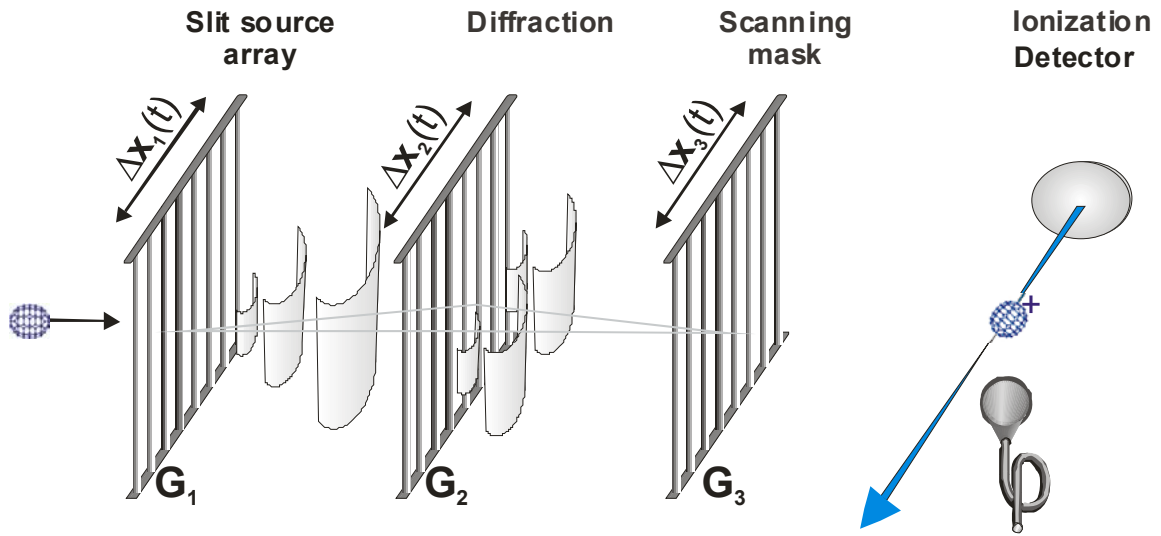

**Figure 1: Interferometer for large molecules.**
Three gratings of equal period are seperated by about the Talbot length. The first grating $G_1$ acts as an incoherent array of slit sources. Each of them illuminates the second grating with a transverse coherence of a few slit distances. Interference then leads to a self-image of the grating $G_2$ at the position of $G_3$. The molecular interferogram is scanned by the third grating. Transmitted molecules are ionized and counted. Inertial forces due to the rotation of the earth, the gravitational acceleration, and acoustic vibrations of the apparatus may lead to a lateral displacement of the interference pattern that depends on the molecular velocity.

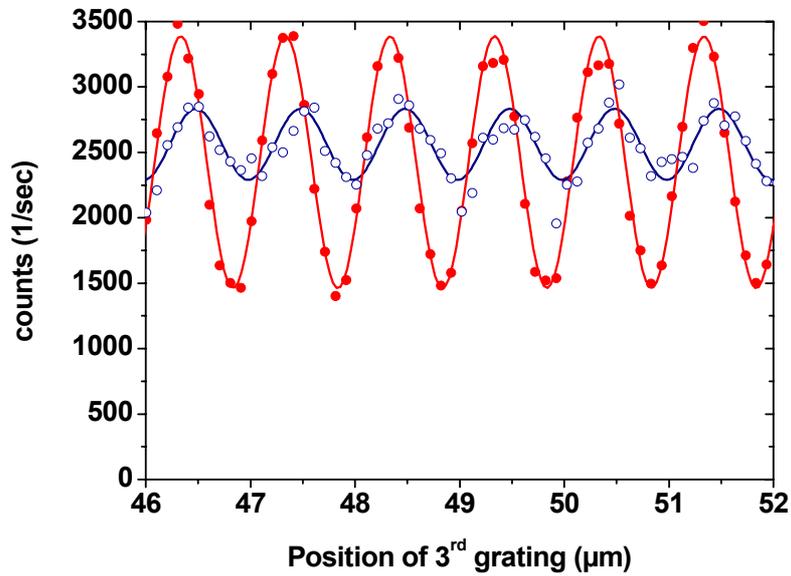

**Figure 2:**

High contrast $C_{70}$ Talbot-Lau interferograms. The high contrast curve (V = 39.5 %, full circles) was recorded while the optical table was inflated. The low contrast curve (V=10.5%, open circles) corresponds to the situation when the table was in direct contact with the floor. Both curves were recorded for molecules at 190 m/s. The solid lines represent a sine-fit with offset to guide the eye. The best possible interference contrast can be even increased to 50 %, when vibrations due to the laser-cooling circuit are suitably decoupled.

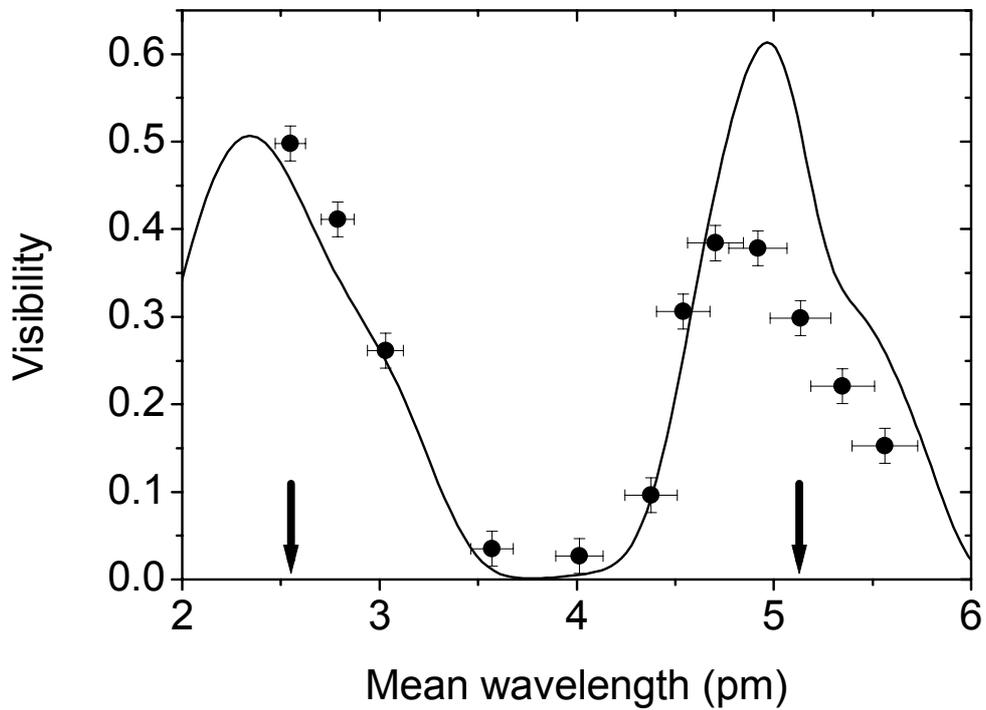

**Figure 3:**

Interference fringe visibility as a function of mean de Broglie wavelength. The circles represent the experimental data. The solid line is derived from a theoretical model which includes all experimental parameters without any further fit. The arrows indicate the Talbot wavelength for the chosen grating separation of L=38 cm. The good agreement between theory and experiment at wavelengths about 2 pm and the discrepancy at about 5 pm is consistent with the assumption of fixed pendulum oscillations of the interferometer (see text) with an amplitude of about 50 nm at a frequency around 100 Hz.

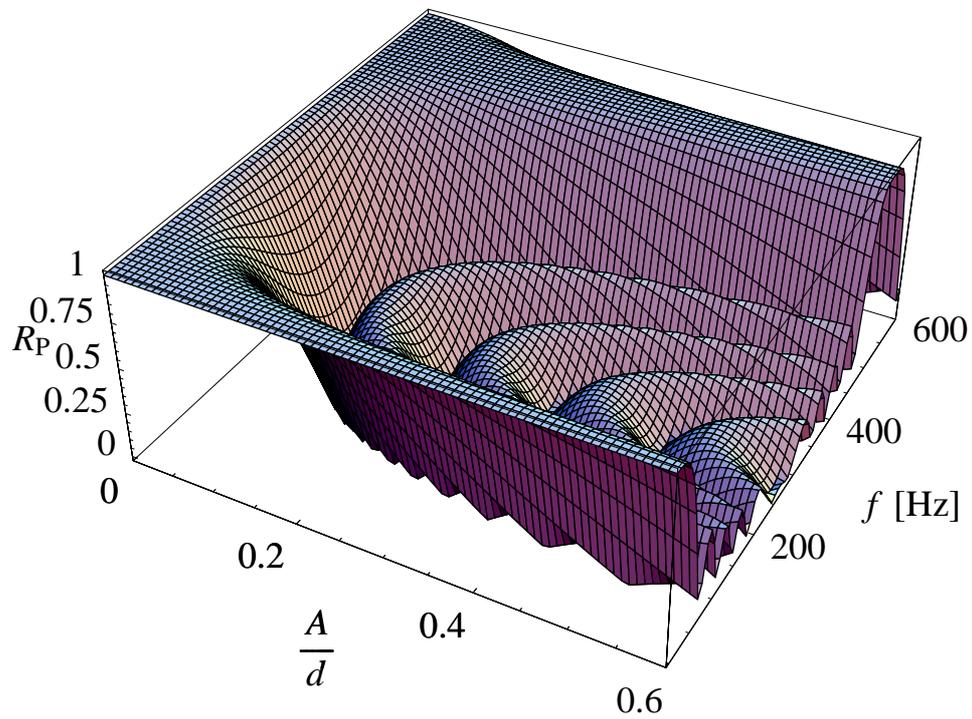

**Figure 4:**

The expected reduction of visibility in the Talbot-Lau setup assuming a common harmonic oscillation of the gratings as a function of the oscillation amplitude and frequency. We choose the parameters of our setup (L=0.38m, $v_z$=200 m/s).

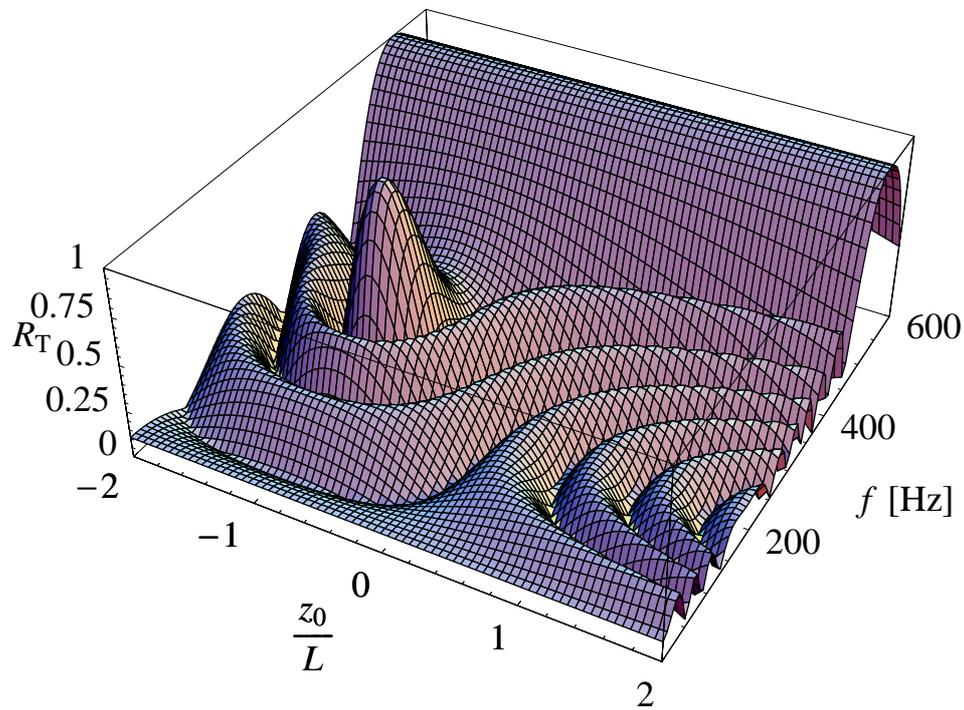

**Figure 5:**

The expected visibility reduction if the gratings perform a harmonic torsional motion about a common rotation axis. The parameter $z_0$ is the longitudinal position of the first grating with respect to the axis of rotation. We choose a maximum angular velocity of $\Omega_0 = 10^{-3}$ rad/s and use the parameters of our setup, L=0.38m, d=1μm, and $v_z$=200 m/s.

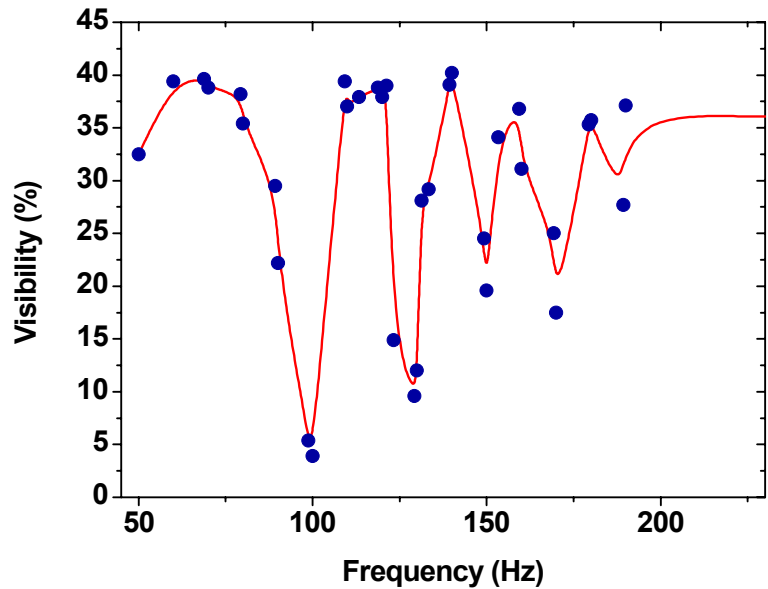

**Figure 6: Influence of acoustic noise on the fringe visibility: frequency dependence**

Interference fringe visibility for $C_{70}$ at a most probable velocity of $v_{mp}$=190 m/s. The frequency of the noise source was varied at a fixed nominal vibration amplitude of 15 nm measured on the outside of a flange on the vacuum chamber. The solid line is a spline curve to guide the eye. Note, that a vibration amplitude of 15 nm in one place can correspond to a much larger amplitude at a different place, outside and inside the vacuum chamber, and that these places will move when the frequency is varied.